
\magnification=1200
\font\titlea=cmb10 scaled\magstep1

\baselineskip=12pt
\rightline{IC/93/81}
\rightline{UG-3/93}
\rightline{hepth/9305039}
\baselineskip=18pt
\vskip .5cm
\centerline{\titlea On the Twisted $N = 2$ Superconformal Structure in }
\centerline{\titlea $2 d$ Gravity Coupled to Matter}
\vskip 2cm
\baselineskip=14pt
\centerline{Sudhakar Panda}
\bigskip
\centerline{\it Institute for Theoretical Physics}
\centerline{\it Nijenborgh 4, 9747 AG Groningen}
\centerline{\it The Netherlands}
\bigskip
\centerline{and}
\bigskip
\centerline{Shibaji Roy}
\bigskip
\centerline{\it International Centre for Theoretical Physics}
\centerline{\it Trieste - Italy}
\vskip 1.0cm
\baselineskip=18pt
\centerline{\titlea Abstract}
\bigskip
It is shown that the two dimensional gravity, described  either
in the conformal gauge (the Liouville theory)
or in the light cone gauge, when coupled to matter possesses an infinite
number of twisted $N = 2$ superconformal symmetries. The central charges of
the $N~=~2$ algebra for the two gauge choices are in general different.
Further, it is argued
that the physical states in the light cone gauge theory can be obtained from
the Liouville theory by a field redefinition.
\vfill
\eject

 In the last few years, two dimensional conformal field theories coupled to
two dimensional gravity has been studied in great detail [1] in two distinct
formulations. In one formulation, known as the matrix models, the string
worldsheet is discretized into many triangles in a careful way and the
summation over all possible triangulations is thus equivalent to the integral
of the metric over all possible geometries [2]. In the other formulation,
known as the continuum approach, the two dimensional metric is fixed by
a suitable gauge and the quantization is performed subsequently. This allows
a choice in fixing the gauge and generally one chooses either the conformal
gauge [3] or the light cone gauge [4]. Though both the
gauge choices gave equivalent results but it was realized that
the conformal gauge is more suitable not only for computational reasons
but also for further developments of the theory. In fact, some of the matrix
model results are obtained directly in the continuum approach following
the conformal gauge choice [5], where the conformal degree of freedom of the
metric is taken as the Liouville field and thus the gravity sector is realized
by the Liouville action.
In a different development, it has been shown that, the 2d topological
gravity coupled to topological matter gives a field theoretic description
of the matrix model formulation of 2d quantum gravity [6]. In fact, it is
proven in ref.[7] that the matrix model formulation of 2d gravity, 2d
topological field theories and the intersection theory of the moduli space
of Riemann surfaces are all equivalent [8].
Although, one still needs to have a complete understanding of the matrix
model results in terms of the continuum approach, yet it might be natural
to expect a topological structure in the continuum approach.
Such a structure is already revealed in the continuum approach [9], again with
the conformal gauge choice of the metric. More recently, it has been
shown that almost all string theories, including the bosonic string, the
superstring and W-string theories, possess a topologically twisted $N = 2$
superconformal symmetry [10] which is a signal that there might be a
connection between topological theories and the field theoretical approach
of gravity coupled to matter system.

As mentioned above, in the continuum approach, 2d gravity can be treated in
the light cone gauge where the the metric degrees of freedom are fixed by
$h_{+-}~=~h_{-+}~=~1/2$ and $h_{--}~=~0$ [4]. It was argued there that the
renormalizability of the theory can be best understood in this gauge. A
remarkable feature of this theory is the presence of an unexpected $SL(2,R)$
current algebra which is responsible for the complete solvability of the
system. As a
technical interest, it is desirable to examine if this formulation of 2d
gravity coupled to conformal matter also possesses the above topologically
twisted $N = 2$ superconformal symmetry. In this letter, we show that this
is indeed true. Furthermore, we indicate briefly how the physical states
in this theory are same as those in the conformal gauge (including the
discrete states) up to a redefinition of the fields.

Let us recapitulate briefly how the twisted $N = 2$ superconformal symmetry
arises in the conformal gauge gravity coupled to matter. The $(p,q)$ minimal
models ($gcd (p,q) = 1$) coupled to the Liouville field
can be described in terms of Coulomb gas representation where the energy
momentum tensors for the matter and Liouville sector are given as
$$
\eqalign{T_M(z)~=&-{1\over 2}~:~\partial X~\partial X~:~+~i Q_M \partial^2 X\cr
         T_L(z)~=&-{1\over 2}~:~\partial \phi~\partial \phi ~:~+~i Q_L
\partial^2 \phi\cr}
\eqno(1)
$$
where $X$ and $\phi$ represent the matter and Liouville fields respectively,
whereas $2Q_M$ and $2Q_L$ denote the background charges. Since, the total
central charge of the combined matter and Liouville system should add up to
26 and the matter sector is characterized by the Virasoro central charge
$ 1 - {6(p-q)^2\over pq}$, it, therefore, follows that
$$
\eqalign{2Q_M&=~\sqrt{{2p\over q}}- \sqrt{{2q\over p}}~=~(\alpha_+~+\alpha_-)
\cr
         2Q_L&=~\pm i (\sqrt{{2p\over q}}+ \sqrt{{2q\over p}})~=~(\beta_+~
+\beta_-)\cr}
\eqno(2)
$$
In the BRST quantization scheme, the BRST current for this system is :
$$
J_B (z)~=~:c (z) [ T_M (z)~+~T_L (z)~+~{1\over 2} T^{bc} (z) ]:
\eqno(3)
$$
where $T^{bc}$ is the energy momentum tensor for the reparametrization
ghost system, consisting of
ghost field $c (z)$ and the anti-ghost field $b (z)$ with conformal weight
$-1$ and 2 respectively and is given by
$$
T^{bc} (z)~=~- 2 :b (z) \partial c(z):~-~:\partial b (z) c (z):
\eqno(4)
$$
It has been noted before that the generators
$$
\eqalign{T (z)&=~T_M (z)~+~T_L (z)~+~T^{bc} (z)\cr
         G^+ (z)&=~J_B (z)\cr
         G^- (z)&=~b (z)\cr
         J (z)&=~:c (z) b (z):\cr}
\eqno(5)
$$
satisfy an almost topological $N =2$ superconformal algebra; however, with
the above choice of the generators, the algebra does not close but produce
two new fields $c (z)$ and $c \partial c(z)$ [11] which can be seen from
the following OPEs
$$
G^+ (z)~G^+ (w)\sim -10 {c \partial c (w)\over (z-w)^3}~- 5 {\partial ( c
\partial c) (w)\over (z-w)^2}~- {3\over 2} {\partial^2 (c \partial c) (w)
\over (z-w)}
\eqno(6)
$$
$$
J (z)~G^+ (w)\sim {G^+ (w)\over (z-w)}~-~{\partial c (w)\over (z-w)^2}
{}~+~{c (w)\over (z-w)^3}
\eqno(7)
$$
It has been found in refs [9,10] that it is possible to modify the BRST current
in (3) and the ghost number current $J (z)$ in (5) by adding total derivative
terms (it does not affect the BRST charge) in such a way that the modified
generators would  form a closed $N =2$ algebra. To be precise, taking the
modified generators as
$$
\eqalign{G^+ (z)&=~J_B (z)~+ a_1~ \partial (c \partial \phi)(z)~+a_2~\partial
(c \partial X) (z)~+a_3~ \partial^2 c (z)\cr J (z)&=~:c (z)b (z):~+~
a_4~\partial \phi (z) ~+a_5~\partial X (z)\cr}
\eqno(8)
$$
where $a_i~(i = 1,2,3,4,5)$ are arbitrary parameters, we find the following
twisted $N =2$ superconformal algebra:
$$
\eqalign{T (z) T (w) &\sim~{2 T (w)\over (z-w)^2}~+~{\partial T (w)\over
(z-w)}\cr T (z) G^{\pm} (w)&\sim~{{1\over 2}(3\mp1) G^{\pm} (w)\over (z-w)^2}
{}~+~{\partial G^{\pm} (w)\over (z-w)}\cr T (z) J (w)&\sim~{-{1\over 3} c^{N=2}
\over (z-w)^3}~+~{J (w)\over (z-w)^2}~+~{\partial J (w)\over (z-w)}\cr
J(z) J(w)&\sim~{{1\over 3} c^{N=2}\over (z-w)^2}\cr J(z) G^{\pm} (w)&\sim~
\pm{G^{\pm} (w)\over (z-w)}\cr G^+ (z) G^- (w)&\sim~{{1\over 3} c^{N=2}\over
(z-w)^3}~+~{J (w)\over (z-w)^2}~+~{T (w)\over (z-w)}\cr G^{\pm} (z)
G^{\pm} (w)&\sim~0\cr}
\eqno(9)
$$
provided the $a_i$'s satisfy the following conditions:
$$
\eqalign{a_1&+~a_4~=~0\cr
         a_2&+~a_5~=~0\cr
         a_1^2&+~a_2^2~+~2 a_3 - 1~=~0\cr
         2i&Q_M a_2~+~2i Q_L a_1~-~2 a_3~+~3~=~0\cr}
\eqno(10)
$$
and where $c^{N=2}~=~6 a_3$ is the central charge of the corresponding
untwisted
$N =2$ superconformal algebra. Note that here we have three unknown parameters
but only two independent conditions governing them. Thus eqn (10) can be
satisfied in many ways and consequently we have infinite number of $N =2$
algebras (i.e. with different central charges ) as the underlying symmetry of
this theory. In ref.[10], a particular solution to eqn (10) i.e. $a_2~=0$
is chosen so that (taking the `$-$' branch for$Q_L$ in eqn (2)) we have
$a_1~=~-\sqrt{{2q\over p}}$ and $c^{N=2}~= 6 a_3~=~3 (1 - {2q\over p})$.
However, it was pointed out in ref.[12] that there is an ambiguity in
choosing the current $\partial\phi$ for deforming the generators $G^+$ and
$J$ because of the fact that when the cosmological constant in the Liouville
action is taken to be non-zero, the Liouville equation of motion implies that
$\partial \phi$ can not be considered as holomorphic current any more. This
situation, of course, will correspond to the case of putting $a_1~=0$
and we will
be left with only two $N=~2$ superconformal algebras (by the interchange of
$p$ and $q$ every where in the above discussion). However, we will not face
such a problem if we choose the light cone gauge instead of conformal gauge
as we will see below.

As shown in [4], in the light cone gauge, the non-zero component of the
metric $h_{++}$ admits a decomposition in terms of the three generators of
the non-compact group $SL(2,R)$ which satisfy the following current algebra:
$$
j^a (z)~j^b (w)~\sim~{f^{ab}_c~j^c (w)\over (z-w)}~+~{{k\over 2}~\eta^{ab}
\over (z-w)^2}
\eqno(11)
$$
where $a, b~= 0, \pm$ are $SL(2,R)$ indices, $k$ is the level of the current
algebra, $\eta$ is the Killing metric with non-zero components $\eta^{+-}~
=~\eta^{-+}~=~- 2 \eta^{00}~=~2$ and the non-zero structure constants are
given as $f^{0 +}_+~=~- f^{0 -}_-~=~- {1\over 2}~f^{+ -}_0~=~-~1$. The residual
gauge invariance is generated by current $j^+$ and the energy momentum tensor
$T_G$. The latter is given by the modified Sugawara form
$$
T_G (z)~=~{1\over k - 2}~: \eta_{ab}~j^a (z) j^b (z) :~-~\partial j^0 (z)
\eqno(12)
$$
and the associated Virasoro central charge is ${3k\over k-2}~+~6 k$. With
respect to this energy-momentum tensor, the currents $j^+, j^0$ and $j^-$
have conformal dimension $0, 1$ and $2$ respectively which can be seen from
the following operator product expansions:
$$
\eqalign{T_G (z)~j^+ (w)&~\sim~~{\partial j^+ (w)\over (z-w)}\cr
         T_G (z)~j^0 (w)&~\sim~~{- k\over (z-w)^3}~+~{j^0 (w)\over (z-w)^2}
{}~+~{\partial j^0 (w)\over (z-w)}\cr T_G (z)~j^- (w)&\sim~~{2 j^- (w)\over
(z-w)^2}~+~{\partial j^- (w)\over (z-w)}\cr}
\eqno(13)
$$
Including the matter coupling, the total energy-momentum tensor in this case is
$$
T (z)~=~T_G (z)~+~T_M (z)~+~T^{bc} (z)~+~:~\partial \zeta \epsilon (z) :
\eqno (14)
$$
where the extra ghost system $(\zeta, \epsilon)$ having conformal weights
(0,1) and ghost-number ($-1$,1) is the consequence of the symmetry associated
with the generator $j^+$. This extra ghost system has Virasoro central charge
$-2$. Thus, taking the matter system again as the $(p,q)$ minimal conformal
matter, the central charge balance equation, now reads as
$$
{3k\over k-2}~+~6k~+~1~- {6 (p-q)^2\over pq}~-~26~- 2~=~0
\eqno(15)
$$
which admits two solutions for $k$ i.e. either $k~=~{p\over q}~+~2$ or
$k~=~{q\over p}~+~2$. We expect that at these values of the level of the
current algebra, the combined matter and gravity theory to possess twisted
$N~=~2$ superconformal algebra
and we will see below that this is indeed true.

The BRST current for this system is given by [13]
$$
J_B (z)~=~: c (z)~[ T_G (z)~+~T_M (z)~+~{1\over 2}~T^{bc} (z)~+~T^{
\zeta \epsilon} (z)] :~+~: \epsilon (z)~j^+ (z) :
\eqno(16)
$$
with $T^{\epsilon\zeta} (z)~= :(\partial\zeta) \epsilon (z):$ and
$T_G (z), T_M (z)$ as given in (12) and (1) respectively. However, as in
the case of the conformal gauge,
this algebra also does not close. Not only the fields $c$,
$c~\partial c$ but
also a new field $( c\epsilon j^+ )$ is generated in the algebra which can
be seen from the following OPE:
$$
J_B (z)~J_B (w)~\sim~{- 10 c \partial c (w)\over (z-w)^3}~-~{5 \partial
( c \partial c ) (w)\over (z-w)^2}~-~{3\over 2}~{\partial^2 ( c \partial c )
(w)\over (z-w)}~+~{\partial ( c \epsilon j^+ ) (w)\over (z-w)}
\eqno(17)
$$
Nevertheless, as in the previous case, we can define the $N = 2$ generators
by modifying the BRST current as well as the ghost number current to obtain
a closed algebra which is again a topologically twisted $N = 2$ superconformal
algebra. To be definite, we take $T (z)$ as in (14), $G^- (z)~=~b (z)$ and
$$
\eqalign{G^+ (z)&~=~J_B (z)~+~A_1~\partial (c \zeta \epsilon) (z)~+~A_2~
\partial^2 c (z)~+~A_3~\partial (c j^0) (z)~+~A_4~\partial (c\partial X)
(z)\cr J (z)&~=~: c(z) b(z) :~+~A_5~: \epsilon (z) \zeta (z) :~+~A_6~j^0 (z)
{}~+~A_7~\partial X (z)\cr}
\eqno(18)
$$
where $J_B$ is as given in (16). We find that these generators satisfy the
topologically twisted $N =2$ superconformal algebra exactly as given in (9)
with $c^{N=2}~=~6 A_2$ provided the $A_i$'s ($i = 1,2,.....,7$) obey the
following relations:
$$
\eqalign{A_1&~-~A_5~=~0\cr A_3&~+~A_6~=~0\cr A_4&~+~A_7~=~0\cr A_1&~+~A_3
{}~-~1~=~0\cr A_1&+~2 A_2~+~k A_3~-~2 i Q_M A_4~-~3~=~0\cr 2&A_1^2~+~4 A_1
{}~+~4 A_2~+~k A_3 (4 - A_3)~-~2 A_4 (A_4 + 4 i Q_M)~-~10~=~0\cr}
\eqno(19)
$$
Again we notice that there are three independent unknown parameters but
two relations governing them. We can fix two of them in terms of the
third one as follows:
$$
\eqalign{A_1&=~{1\over k-2}~[ 1 \pm \sqrt{ (k-3)^2 - 2 A_4 (k-2) (A_4
+ 2 i Q_M)} ]\cr A_2&=~1 + i Q_M A_4 - {k-1\over 2(k-2)} [ (k-3)\mp
\sqrt{(k-3)^2 - 2 A_4 (k-2) (A_4 + 2iQ_M)} ]\cr}
\eqno(20)
$$
Thus for different values of $A_4$ we have a topologically twisted $N=2$
superconformal algebra with different central charges given by $6 A_2$. In
particular, restricting to $A_4 = 0$ and substituting for $k$ in terms of $p$
and
$q$, as found earlier, we obtain that the central charge of the corresponding
$N=2$ theory is given by
$$
c^{N=2}~=~6 ({p\over q} - {q\over p} + 1)
$$
or 6, which is a particular case  $p = q$ of the above and corresponds to
the case of $c_M = 1$ coupled to gravity, described in the light cone gauge.
Comparing to the corresponding expression for $c^{N=2}$ in the conformal
gauge, which is $c^{N=2}~=~3 (1 - {2q\over p})$, we observe that the
underlying $N =2$ theory have different central charge for the two different
gauge choices of the metric. In fact, this is true for the generic case also.
This in turn implies that unless we can establish an automorphism under which
the generators of the $N=2$ algebra, in these two gauges, have one to one
correspondence and the central charge is same for both the cases, it may be
ambiguous to determine the physical state spectrum of the matter coupled to
gravity theory by relying on the $N = 2$ symmetry alone. We defer
further discussion on this issue for a later occasion and proceed to show
how the physical states, including the discrete states, of this theory
corresponding to the above two gauge choices are related by a
field redefinition.

We recall that the physical states are obtained by studying the cohomology
classes of the BRST charge, which we denote for the two cases as
$Q_B^{Conf,LC}$, defined as
$$
Q_B^{Conf,LC}~=~ \int~dz~J_B^{Conf,LC} (z)
$$
where $J_B^{Conf}, J_B^{LC}$ are the BRST currents and defined in (3) and
(16) respectively. The physical states of the theory are the states
which are in the kernel of the BRST charge modulo its image. The analysis of
the physical states for standard ghost number states have been done long ago
both for conformal gauge [3] and for the light cone gauge [13].
However, in recent years, the discovery of discrete states associated with
different ghost numbers [14] has drawn much attention. By a simple argument,
we will see that there is a one to one correspondence among the elements of
the physical state spaces of the above two cases, as it should be. For this
purpose, we consider the free field realization of the $SL(2,R)$ current
algebra for the level $k$, known as Wakimoto construction [15] and write the
$SL(2,R)$ currents as
$$
\eqalign{j^+ (z)&~=~\beta (z)\cr
j^0 (z)&~=~:\beta (z) \gamma (z):~+~\sqrt{{k-2\over 2}} \partial \phi (z)\cr
j^- (z)&~=~:\beta (z) \gamma^2 (z):~+~2 \sqrt{{k-2\over 2}} \gamma (z)\partial
\phi (z)~+~k \partial\gamma (z)\cr}
\eqno(21)
$$
where $\phi, \beta, \gamma$ are free bosonic fields with
OPE~~$\beta (z)\gamma (w)~\sim~(z-w)^{-1}$,

\noindent $~\phi (z)\phi (w)~\sim~- \log (z-w)$.
In terms of these free fields, the gravitational energy momentum tensor as in
(12) takes the form
$$
T_G (z)~=~- :\partial \beta (z) \gamma (z):~-~{1\over 2} :\partial \phi (z)
\partial \phi (z):~+i Q_L \partial^2 \phi (z)
\eqno(22)
$$
where $Q_L$ is as given in (2). Note that $\beta$ and $\gamma$ have conformal
dimensions 0 and 1 respectively with respect to this $T_G$. It is now clear
that if we identify the Wakimoto field $\phi$ as the free Liouville field,
then we have
$$
T_G (z)~=~T^{\beta\gamma}(z)~+~T_L (z)
\eqno(23)
$$
where $T^{\beta\gamma} (z)~=~- :(\partial \beta (z)) \gamma (z):$.
The BRST current then reads as
$$
J_B^{LC} (z)~=~J_B^{Conf}~+~:c (z) [ T^{\beta \gamma} (z)~+~T^{\epsilon
\zeta} (z)]:~+~:\epsilon (z) \beta (z):
\eqno(24)
$$
Thus we have
$$
Q_B^{LC}~=~Q_B^{Conf}~+~Q_B^{(1)}~+~:[ Q_B^{(1)}, \int dz c(z)\partial
\zeta (z) \gamma (z) ]:
\eqno(25)
$$
where $Q_B^{(1)}~= \int dz \epsilon (z) \beta (z)$. Note that both
$Q_B^{Conf}$ and $Q_B^{(1)}$ are independently nilpotent and anticommute with
each other. Now from the knowledge of the physical state space which are in
the cohomology class of $Q_B^{Conf}$, we can derive the states in the
cohomology class of $Q_B^{LC}$ as follows. Let $|\psi>^{Conf}$ be
a physical state
in the conformal gauge i.e.
$$
Q_B^{Conf}~|\psi>^{Conf}~=~0
\eqno(26)
$$
The solution to this equation can be written symbolically as
$$
|\psi>^{Conf}~=~{\cal P} (\partial X,\partial \phi, b, c )
{}~V (X)~V (\phi)~|0>
\eqno(27)
$$
where ${\cal P}$ is a differential polynomial, $V$'s are vertex operators
and $|0>$ is the $SL(2,C)$ vacuum.
Introducing an unitary operator $U$ as
$$
U~=~e^{-\int dz c (z) \gamma (z) \partial \zeta (z)}
\eqno (28)
$$
we find a relation between the BRST charges in the conformal and the light
cone gauges as follows,
$$
Q_B^{LC}~=~U~ [ Q_B^{Conf} + \int dz \epsilon (z) \beta (z) ]~ U^{-1}\eqno (29)
$$
The physical states in the light cone gauge will be rotated accordingly,
$$
|\psi>^{LC}~=~U |\psi>^{Conf}\eqno(30)
$$
The effect of $U$ on $|\psi>^{Conf}$ is just to shift the field $b$ by
$b+\gamma\partial\zeta$. To be explicit, let us recall [16] that the
physical state spectrum of the 2d gravity in the conformal gauge coupled to
$(p,q)$ minimal matter are generated by three types of operators $x$, $y$
and $w(w^{-1})$. Here $x$, $y$ are the spin zero, ghost number zero
operators and are called the ground ring generators, whereas $w(w^{-1})$ has
ghost number $-1(+1)$. Any physical operator at ghost number $-n$ can be
written as
$$
{\cal O}_{n,i,j}~=~ w^n x^i y^j\eqno(31)
$$
where $i$, $j$ are integers with the restriction $0\le i\le p-2$,
$0\le j\le q-2$. The ground ring generators in the light cone gauge
take the form
$$
\eqalign{ x &= [b c + \gamma\partial \zeta c + {3\over 4} {\sqrt {2q\over p}}
(i\partial X + \partial \phi)] e^{i\alpha_{1,2}X + i\beta_{1,2}\phi}\cr
y &= [b c + \gamma \partial \zeta c - {3\over 4} {\sqrt {2p\over q}}
(i\partial X - \partial \phi)] e^{i\alpha_{2,1} X + i\beta_{2,1}\phi}\cr}
\eqno(32)
$$
where $\alpha_{m,m'} = {1\over 2}[(1-m)\alpha_+ + (1-m')\alpha_-]$ and
$\beta_{n,n'} = {1\over 2}[(1-n)\beta_+ + (1-n')\beta_-]$. The other
generators $w$, $w^{-1}$ with ghost number $-1$, +1 in general would have
the form
$$
\eqalign {w &= {\cal P} (\partial X, \partial \phi, b + \gamma \partial\zeta,
c) e^{i\alpha_{q-1,
1} X + i\beta_{1, p+1}\phi}\cr
w^{-1} &= c e^{i\alpha_{q-1,1} X + i\beta_{-q+1,1}\phi}\cr}\eqno(33a)
$$
or
$$
\eqalign{w &= {\cal P} (\partial X, \partial \phi, b +\gamma\partial\zeta, c)
e^{i\alpha_{1,p-1}
X + i\beta_{q+1,1}\phi}\cr
w^{-1} &= c e^{i\alpha_{1,p-1} X + i\beta_{1,-p+1}\phi}\cr}\eqno(33b)
$$
where ${\cal P}$ is a differential polynomial of conformal weight $(p+q-1)$
and ghost number $-1$. We have two sets of $w(w^{-1})$ because we note that
since $w\cdot w^{-1} \sim I$ their multiplication is well defined if we take
$w$ from (33a) and $w^{-1}$ from (33b) or vice versa. Since, for general
$(p,q)$ model the form of $w$ is quite complicated we give its form
for (3,2) model which is pure Liouville gravity
$$
\eqalign{ w &= ({1\over 2} \partial^2 b + {1\over 2} \partial^2 \gamma
\partial \zeta + \partial \gamma \partial^2 \zeta + {1\over 2} \gamma
\partial^3\zeta - 3\partial b b c + 3 \partial b c \gamma \partial \zeta\cr
&-3 b c \partial \gamma\partial \zeta -3 b c \gamma \partial^2\zeta +
3c \gamma^2 \partial \zeta \partial^2 \zeta - {{\sqrt 3}\over 2} \partial b
\partial \phi - {{\sqrt 3} \over 2} \partial \gamma \partial \zeta
\partial \phi\cr
&- {{\sqrt 3} \over 2}\gamma \partial^2\zeta\partial \phi + {\sqrt 3} b
\partial^2\phi + {\sqrt 3}\gamma \partial\zeta \partial^2\phi) e^{{\sqrt 3}
\phi}\cr}\eqno(34)
$$
It is now a simple exercise to check that the operators in (32) and
(34) belong to the relative cohomology of the full BRST charge
given in (25). The Physical states for $c_M = 1$ matter coupled to
2d light cone gauge gravity have been discussed in a recent paper [17].
Following closely the analysis of ref.[18], it has been found there, that
the oscillator part of the physical operators gets a shift $b + \gamma
\partial\zeta$ in place of $b$. This is precisely the field redefinition
we obtain for $c_M < 1$ comparing the BRST charges in two different
gauges.

To conclude, we have shown here that the two dimensional gravity, described
by either the Liouville theory or in the light cone gauge, when coupled to
conformal matter, possesses an infinite number of topologically twisted
$N=2$ superconformal symmetries. The topological central charge for the
two gauge choices are found not to be the same. This indicates that the
analysis of physical states by using the underlying $N=2$ symmetry may
not be unique. By performing a rotation we also argued that the physical
states of the 2d light
cone gauge gravity are same as the physical states found in the conformal
gauge after a shift of the anti ghost field. Doing this rotation in the
opposite way, we notice that the physical states and the BRST operators
for the two cases are same if we ignore the $\beta ,\gamma ,\epsilon ,\zeta$
degrees of freedom. Though
there is no a priori reason, but if we use this information in the analysis
of N = 2 algebra (with Wakimoto realization), we observe that the generators
and the central charge for both the gauge choices are in one to one
correspondece with the Wakimoto field $\phi$ being identified as the
Liouville field.
\vfil
\eject
\noindent{\bf Acknowledgments}

 One of us (S.R.) would like to thank S. Ghosh and M. H. Sarmadi for valuable
 discussions. He also thanks Prof. A. Salam, IAEA, UNESCO and ICTP, Trieste,
 for support.
 S.P.'s work is performed as
part of the research program of the ``Stichting voor Fundamenteel Onderzoek
der Materie'' (FOM).

{\titlea References}
\bigskip
\item{1.} For a review article, see P. Ginsparg and G. Moore, proc. of
TASI 92, ed. J. Harvey and J. Polchinski.
\item{2.} P. Ginsparg, in proc. of Trieste Summer School 91 and references
therein.
\item{3.} F. David, Mod. Phys. Lett. A3 (1988) 1651.
 J. Distler and H. Kawai, Nucl. Phys. B321 (1989) 509.
\item{4.} A. M. Polyakov, Mod. Phys. Lett. A2 (1987) 893.
V.G. Knizhnik, A.M. polyakov and A.B. Zamolodchikov, Mod. Phys. Lett. A3
(1988) 819.
\item{5.} M. Goulian and M. Li, Phys. Rev. Lett. 66 (1991) 2051;
P. DiFrancesco and D. Kutasov, Phys. Lett. B261 (1991) 385;
M. Bershadsky and I. Klebanov, Nucl. Phys. B360 (1991) 559.
\item{6.} K. Li, Nucl. Phys. B354 (1991) 711,725.
\item{7.} E. Witten, Nucl. Phys. B340 (1990) 281; Surv. Diff. Geom.
1 (1991) 243.
\item{8.} See also R. Dijkgraaf in proc. of Cargese Summer School 1991
and references therein.
\item{9.} B. Gato-Rivera and A. Semikhatov, Phys. Lett. 288B (1992) 295.
\item{10.} M. Bershadsky, W. Lerche, D. Nemeschansky and W. Warner,
HUTP-A034/92
\item{11.} R. Dijkgraaf, E. Verlinde and H. Verlinde, PUPT-1217, IASSNS-HEP-
90/80
\item{12.} S. Mukhi and C. Vafa, HUTP-93/A002, TIFR/TH/93-01.
\item{13.} T. Kuramoto, Phys. Lett. 233B (1989) 363; Z. Horvath, L. Palla and
P. Vecsernyes, Int. J. Mod. Phys. A4 (1989) 5261; K. Itoh, Nucl. Phys. B342
(1990) 449.
\item{14.} B. Lian and G. Zuckerman, Phys. Lett. 254B (1991) 417; Comm. Math.
Phys. 135 (1991) 547.
\item{15.} M. Wakimoto, Comm. Math. Phys. 104 (1986) 605.
\item{16.} H. Kanno and M. H. Sarmadi, preprint IC/92/150 (1992); S. Panda
and S. Roy, preprint UG-1/93, IC/93/13 (to appear in Phys. Lett. B).
\item{17.} N. Marcus and Y. Oz, Nucl. Phys. B392 (1993) 281.
\item{18.} P. Bouwknegt, J. McCarthy and K. Pilch, Comm. Math. Phys.
145 (1992) 541.

\bye